\documentclass[aps,twocolumn,groupedaddress,amsmath,ams-fonts,showpacs]{revtex4}
\usepackage[dvipdfmx]{graphicx}
\usepackage{dcolumn}
\usepackage{bm}
\usepackage{here}
\usepackage{relsize}
\usepackage{amssymb}
\usepackage{txfonts}
\usepackage{bm}
 \usepackage{braket}
\allowdisplaybreaks[4]
\usepackage{color}
 \usepackage{braket}

\usepackage[pdftex]{hyperref}
\def\[#1\]{\begin{align}#1\end{align}}
\def\(#1\){\begin{itemize}#1\end{itemize}}

\begin{document}

\title{
Quantum skyrmions in two-dimensional chiral magnets
}
\author{Rina Takashima$^{1}$}
\email{takashima@scphys.kyoto-u.ac.jp}
\author{Hiroaki Ishizuka$^{2}$}
\author{Leon Balents$^{3}$}
\affiliation{$^1$Department of Physics, Kyoto University, Kyoto 606-8502, Japan}
\affiliation{$^2$Department of of Applied Physics, The University of Tokyo, Tokyo 113-8656, Japan }
\affiliation{$^3$Kavli Institute for Theoretical Physics, University of California, Santa Barbara, CA 93106, USA }
\date{\today}
\begin{abstract}
We study the quantum mechanics of magnetic skyrmions in the vicinity of the skyrmion-crystal to ferromagnet phase boundary in two-dimensional magnets. We show that the skyrmion excitation has an energy dispersion that splits into multiple bands due to the combination of magnus force and the underlying lattice. Condensation of the skyrmions can give rise to an intermediate phase between the skyrmion crystal and ferromagnet: a quantum liquid, in which skyrmions are not spatially localized.   We show that the critical behavior depends on the spin size $S$ and the topological number of the skyrmion.   Experimental signatures of quantum skyrmions in inelastic neutron-scattering measurements are also discussed. 
\end{abstract}
\maketitle
\section{Introduction} 
A magnetic skyrmion is a localized spin texture characterized by an integer topological index called the skyrmion number:
\[
\mathcal N=\frac{1}{4\pi} \int d^2 r \ \bm n(\bm r)\cdot [\partial_x \bm n(\bm r) \times \partial_y \bm n(\bm r)],\label{skyrmion_num}
\]
where $\bm n(\bm r)$ is a unit vector which describes the magnetization density. 
Skyrmions were predicted theoretically in chiral magnets decades ago\cite{Bogdanov1994}, and recently have been discovered experimentally\cite{Rossler2006, Muhlbauer2009,Yu2010a}. They have attracted broad interest due to the extraordinary properties.\cite{Kanazawa2011,Schulz2012,Mochizuki2012, Romming2013, Sampaio2013, Nagaosa2013} Typically they appear in a periodic skyrmion crystal (SkX) phase, examples of which are being continuously discovered. Some specific properties of the skyrmions, such as their size, vary from system to system. For instance, the size of skyrmions in Fe$_{0.5}$Co$_{0.5}$Si thin films is $\sim 90$ nm ,\cite{Yu2010a}\
while skyrmions on Fe thin film deposited on the Ir surface are $\sim 1$ nm.\cite{Heinze2011} In addition, possible realizations of SkX phases in classical frustrated magnets have been studied theoretically \cite{Okubo2012,Leonov2015, Lin2015}; their size is comparable to the lattice spacing.   

In thin films of Fe$_{0.5}$Co$_{0.5}$Si, the SkX and adjacent phases have been observed by real-space imaging in the presence of a magnetic field~\cite{Yu2010a}. In the high field region just below the phase boundary between the
field-induced ferromagnetic (FM) phase and the SkX phase, a small
density of skyrmions are introduced in equilibrium in the FM state. On reducing the field, the number of skyrmions increases, and they eventually form a dense crystal. It is known that skyrmions in thin films including this material remain stable in the wide range of temperature from the order of $\sim100$K\cite{Yu2011} to nearly zero temperature\cite{Yu2010a, Tonomura2012}. 

In the work discussed above, a classical description of the spins is sufficient, and skyrmions may simply be regarded as textures of quasi-static vector moments.  Here, we pursue the possibility of observing skyrmions in the quantum realm.  Several basic theoretical questions arise: What are the quantum states of a skyrmion?  What is the corresponding spectrum, and what are the quantum numbers associated with a skyrmion?  Is a skyrmion a quasiparticle?   How do quantum effects change universal properties of skyrmion systems?  Are there new skyrmion phases besides the SkX one induced by quantum dynamics? Practically, we would also like to understand observable consequences of quantum skyrmions, and when they should be visible.  Clearly,  quantum effects are strongest when the spin quantum number $S$ and the skyrmion diameter $L_s$ are small.  One would like to understand their magnitude, and how it scales with these parameters.  We address these questions and provide some answers in this paper.  

\begin{figure}[t]
\includegraphics[clip, width=0.8\columnwidth]{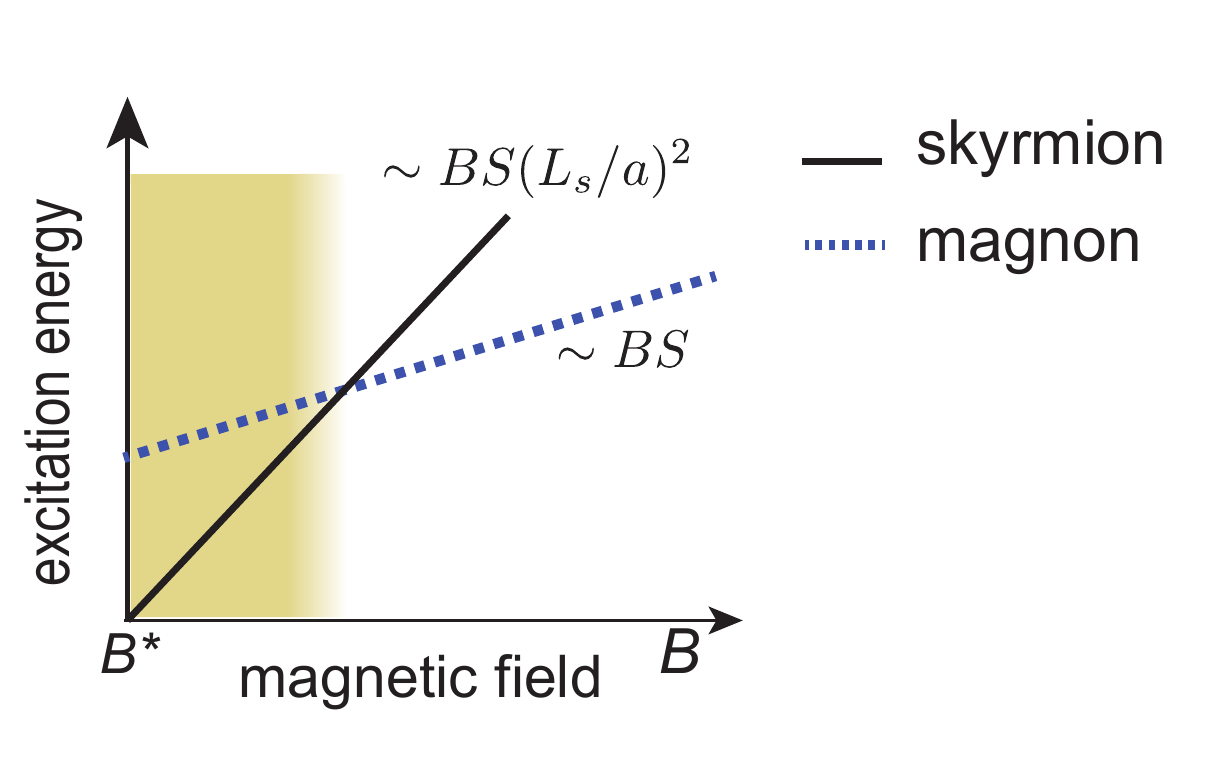}
\caption{
Schematic field dependence of the excitation energy. 
 $L_s$ is the characteristic length of skyrmions, and $a$ is the lattice spacing. Our low-energy effective theory focuses on the colored region, where a skyrmion is the lowest excitation. See Sec.~\ref{sec:action} for details.  }
\label{field_dep}
\end{figure}

First, we argue that a skyrmion indeed becomes a quasiparticle in the appropriate quantum regime.  This occurs close to the phase boundary between the FM and SkX phases, at zero temperature.  Here, for systems with large skyrmions such as Fe$_{0.5}$Co$_{0.5}$Si, the FM/SkX transition is understood to be of the so-called ``nucleation'' type~\cite{Gennes1975,Robler2011}; it is not associated with a small order parameter. 
This implies that a small deviation of the order parameter, like a magnon, costs more energy than the excitation of a skyrmion, at least close to the phase boundary of the FM state (Fig.~\ref{field_dep}).  This allows, in a quantum description, the skyrmion to become a stable quasiparticle.  In this limit, we show that a skyrmion excitation has a non-trivial band structure, which depends on the spin $S$ and the skyrmion number $\mathcal N$.  This structure is the quantum descendent of the well-known Magnus force dynamics of classical skyrmions.  The skyrmion states can be probed by inelastic neutron-scattering within the FM phase.  They are clearly distinct from magnons in possessing many bands (while a FM state has a single magnon branch), with particle dispersion and spectral weight characteristic of skyrmions.  

Next, we consider how quantum  dynamics modifies the classical nucleation transition. The nucleation scenario works well if skyrmions are fully classical, in other words, the size of a skyrmion is much larger than the lattice spacing of the underlying lattice. However, the fate of this transition is nontrivial when the skyrmion develops a quantum nature, which can be prominent for small skyrmions, especially when the density is low enough so that they do not form a classical crystal. We find that skyrmions can undergo a variant of {\em Bose condensation} to form a quantum liquid phase between the FM and SkX phases.   We also show that this phase is connected to the FM phase via a phase transition or a crossover depending on the parity of $2S \mathcal N$.

\begin{figure}[t]
\includegraphics[clip, width=0.85\columnwidth]{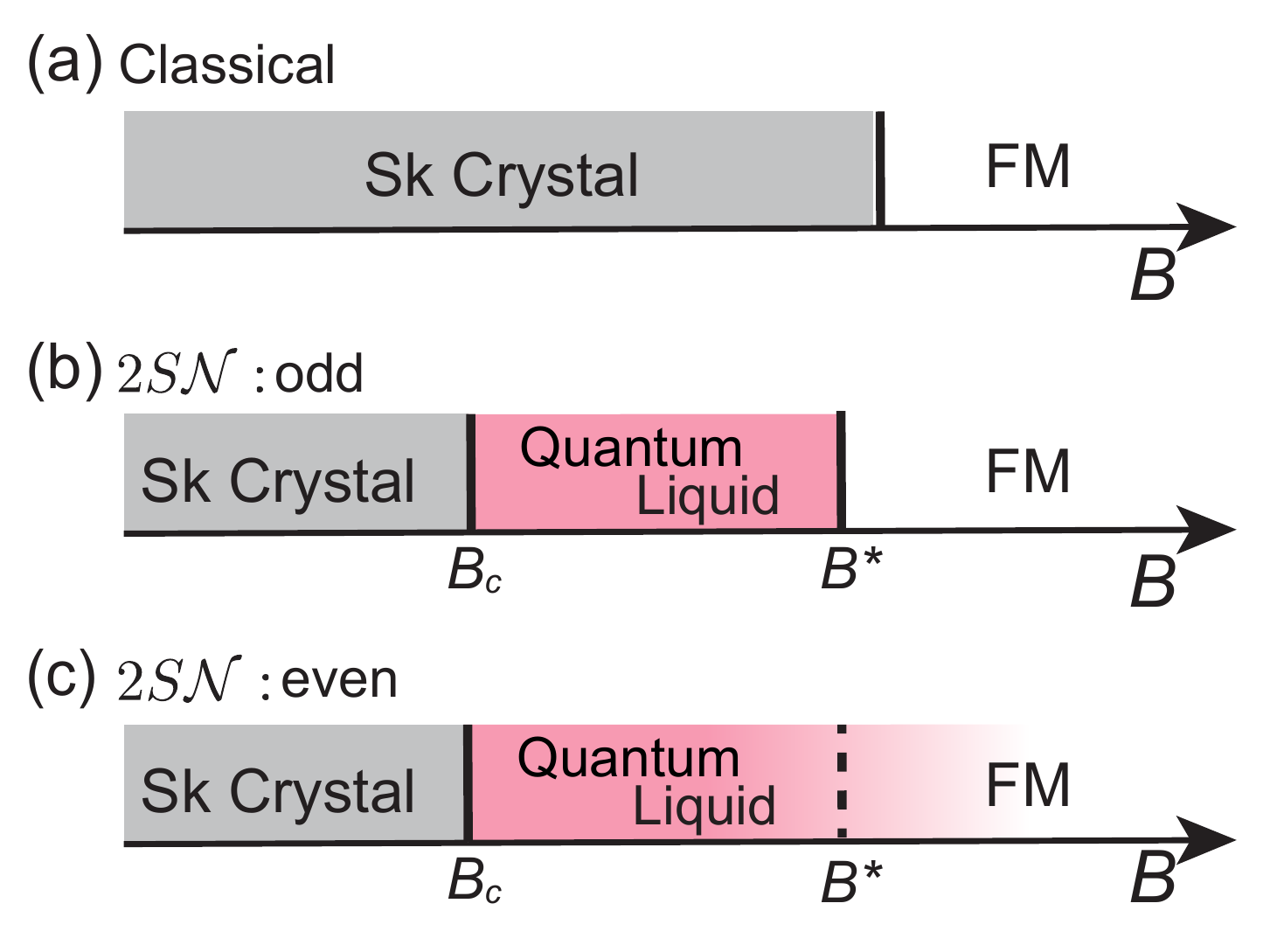}
\caption{
Schematic phase diagrams at zero temperature for (a) classical spins\cite{Robler2011}, (b) quantum spins with even $2S\mathcal N$, and (c) quantum spins with odd $2S\mathcal N$. $S$ is the spin size and $\mathcal N$ is the skyrmion number of a single skyrmion. In (a), skyrmions (Sk) are introduced in the ferromagnetic (FM) background below the critical magnetic field. In (b) and (c), the quantum liquid phase of skyrmions appears. It is connected to the FM phase via the continuous phase transition for (b) or the crossover for (c).    
}
\label{phase}
\end{figure} 

The organization of this paper is as follows. In Sec.~\ref{sec:action}, starting from a two-dimensional spin model on a lattice, we derive the low-energy effective action for single skyrmion excitations. We include the effect of the underlying lattice, which cannot be neglected for small skyrmions. 
Then, from the effective action, we calculate the energy dispersion of the low energy excitation. Section~\ref{intermediate} discusses the phase diagram (Fig.~\ref{phase}) and the critical phenomena in light of the skyrmion spectrum. In Sec.~\ref{neutron}, we show how the quantum states of a skyrmion can be detected through neutron-scattering experiments. Finally, in Sec.~\ref{conclusion}, we summarize our results and discuss the generalizations of our theory. 

\section{Single particle excitation }\label{sec:action}
\subsection{Effective Action}

We first introduce the effective action for a single skyrmion excitation that includes the effects of the underlying crystal lattice.  

We start with the quantum counterpart of an effective Hamiltonian for two dimensional chiral magnets:~\cite{Butenko2010, Robler2011} 
\[
\hat{H}=&-J\sum_{\langle i,j\rangle} \hat{\bm S}_i \cdot \hat{\bm S}_j + D \sum_{i}\sum_{\mu=x,y} \bm e_\mu\cdot \left( \hat{ \bm S }_i \times \hat{\bm S}_{i+\bm e_\mu} \right) \nonumber\\
 &\hspace{40pt}-B\sum_i \hat{S}_i^z-K \sum_i (\hat{S}_i^{z})^2.
\]
Here, the first term is the ferromagnetic ($J>0$) exchange coupling between the nearest neighbor spins, taken for simplicity on a square lattice, and the second term is the Dzyaloshinskii-Moriya (DM) coupling.  $B$ is the external magnetic field perpendicular to the plane, and the last term is an uniaxial anisotropy.

In this paper, we focus on the vicinity of the phase boundary between the FM and SkX phase. While a skyrmion has a large energy under a very high field as $\sim  B S (L_s/a)^2 $, with the size of a skyrmion $L_s$ and the lattice spacing $a$, it has lower energy than a magnon close to the phase boundary of the SkX phase (Fig.~\ref{field_dep}). We thus only consider a skyrmion excitation as the low energy excitation. 

To obtain the effective action for a single skyrmion, we use the path integral formalism. At each site labeled by $i$, the state $ |\bm n_i\rangle $ is parametrized by a unit vector $\bm n_i =(\sin \theta_i \cos \phi_i, \sin \theta_i \sin \phi_i, \cos \theta_i)$ to give $\langle \bm n_i| \hat{\bm S_i} |\bm n_i\rangle =S \bm n_i$. The whole system is represented by a product state of all the sites: $|\Psi(\tau)\rangle = \otimes_{i} |\bm n_i(\tau)\rangle $. 
The distribution function and the action are given by,
\[
Z&=\int  \prod_{i} \mathcal D\bm n_i(\tau) \exp(- \mathcal S[\{\bm n_i \}]),\\
\mathcal S(\{\bm n_i\}) &=i S \int d\tau  \sum_i \dot{\bm n}_i \cdot \bm A(\bm n_i ) +\int d\tau H(\{\bm n_i\}). \label{action} 
\]
The first term in the action is the Berry phase term, where the explicit form of $\bm A(\bm n_i )$ depends on the gauge choice, e.g. $ \dot{\bm n}_i \cdot \bm A(\bm n_i )= \dot \phi_i (1-\cos \theta_i)$ for a certain gauge.  
The second term is ${\rm H}(\{\bm n_i\})  = \langle \Psi (\tau)| \hat H |\Psi (\tau)\rangle $.  

To study the skyrmion excitations, we consider a single skyrmion configuration of ${\bm n}_i$ : $\bm n_i=\bm n_{sk} (\bm r_i - \bm R(\tau)) $. $\bm n_{sk}(\bm r) $ is the unique continuous $O(3)$ vector field with a skyrmion centered at the origin that minimizes the classical energy in the continuum limit. While it cannot be obtained explicitly, $\bm n_{sk}(\bm r)$ is fully specified in this way and we can use it successfully. $\bm R(\tau)=(X(\tau),Y(\tau))$ is the position of the skyrmion. 
 
The resulting effective action for a single skyrmion is given by
\[
 &\mathcal S_{\text{eff}}(\bm R) \nonumber\\
&=\int d \tau \left[\frac{2\pi i S \mathcal N}{a^2}  ( Y\dot{X}-X\dot{Y}) + g\left(\cos \left( \frac{2\pi X}{a} \right)+\cos \left( \frac{2\pi Y}{a} \right)\right) \right], \label{effS}
\] 
where $\dot X$ ($\dot Y$) is the imaginary-time derivative of $X$ ($Y$), $g$ is the magnitude of the periodic potential that arises from the underlying lattice, and $\mathcal N$ is the skyrmion number defined by Eq.~\eqref{skyrmion_num}. The time derivative term arises from the Berry phase term, and has been derived by many authors.\cite{Stone1996, Jonietz2010,Yu2012a}\ It makes $X$ and $Y$ canonically conjugate, and yields the Magnus force in the classical motion of a skyrmion.  

We obtain the periodic potential by considering the leading corrections to the continuum limit beginning with the lattice model. Under some assumptions, the magnitude is  
\[
g&\sim  \frac{2C}{\pi} \varepsilon L_s^2  \exp\left(-C \frac{L_s^2 }{a^2} \right) , \label{g_order}
\]
where ${\varepsilon} \  (\sim JS^2/a^2)$ is the ``energy density'' of a skyrmion in the continuous limit, and  $C \sim  O (1)$ is a dimensionless constant. Details of the derivation are shown in Appendix \ref{ap:eff_S}. 

\subsection{Single Particle Energy}\label{energy}

From the effective action in Eq.~\eqref{effS}, we obtain the effective Hamiltonian of the skyrmion via canonical quantization:
\[
\hat{H}_{\text{eff}}&= g\left(\cos \left( \frac{2\pi \hat X}{a} \right)+\cos \left( \frac{2\pi \hat Y}{a} \right)\right), \\
[\hat{X},\hat{Y}]&=\frac{ia^2}{4 \pi S \mathcal N}\equiv\frac{i a^2}{2 \pi {\rm p}}, \label{commute}
\]
where $2SN_{sk}\equiv  {\rm p} \in \mathbb Z$. The two position operators of the skyrmion, $X$ and $Y$, become commutative in the classical limit, $S\rightarrow \infty$, or in the large skyrmion limit, $a/L_s \to 0$. 

To calculate the eigenstate, we first introduce the translation operators $T_1\equiv e^{-2\pi i \hat{Y}/a  } $ and $T_2 \equiv e^{ 2\pi i \hat {X}/a }$. These operators shift the position of a skyrmion by 
\[
T_1^\dag \hat{ X} T_1 =\hat{ X}+\frac{a}{ {\rm p}},\\
T_2^\dag \hat { Y} T_2 = \hat {Y}+\frac{a }{ {\rm p}}, 
\]
due to the commutation relation in Eq.~\eqref{commute}. These operators are non-commutative, $T_1 T_2 =\exp\left( -2 \pi i /  {\rm p} \right)T_2T_1$, but $T_1$ commutes with $T_2^ {\rm p}$ (translation over the lattice spacing) and vice versa.  In particular this implies further that $T_1^{\rm p}$ and $T_2^{\rm p}$ commute, which is a mathematical expression of the fact that the effective flux per unit cell is an integer multiple (${\rm p}$) of the flux quantum.  Using these operators, the Hamiltonian is given by
\[
\hat H_{\rm eff}&=\frac{g}{2} \left( T_1+T_2\right)+\text{h.c.}. 
\]
To calculate the energy eigenstates, it is convenient to use a basis given by simultaneous eigenstates of $T_1^ {\rm p}$ and $T_2$, 
\[
T_1^ {\rm p} | k_x, k_y \rangle &=e^{i  k_x a} |k_x, k_y\rangle, \\  
T_2 | k_x, k_y \rangle &=e^{i k_y a / {\rm p}} |k_x, k_y\rangle,
\]
with $|k_x| \leq  \pi/a$ and $|k_y|\leq {|\rm p|}\,\pi/a$. We choose a boundary condition such that the operation of $T_1$ on $|k_x, k_y \rangle$ yields 
\[
T_1|k_x, k_y \rangle &= e^{i k_x a/ {\rm p}} |k_x, k_y+2 \pi /a \rangle,
\]
which implies that $|k_x , k_y+2 \pi  {\rm p}/a  \rangle =| k_x, k_y  \rangle $. We also define $|k_x +2 \pi/a, k_y  \rangle =| k_x, k_y  \rangle $.

In this basis, the Hamiltonian is
\[
\hat{H}_{\rm eff} &=g\sum_{k_x, k_y}  \cos \left( \frac{ k_y a}{  {\rm p}} \right) |k_x, k_y  \rangle    \langle k_x, k_y |\nonumber\\
&+ \frac{g}{2}\sum_{k_x, k_y}  \left(   e^{ i k_x a/ {\rm p} } |k_x, k_y+2 \pi/a  \rangle \langle k_x, k_y |+ \text{h.c.} \right) , \label{harper}
\]
which is the same form as Harper's equation.~\cite{Harper1955} There are $ {|\rm p|}=2S|\mathcal N|$ split bands, and the eigenstates are given by
\[
|\psi_{\alpha \bm k}\rangle=\sum_{\ell =0}^{ {\rm p}-1} c_{\alpha, \bm k}(\ell) | k_x,k_y +2 \pi \ell /a \rangle, \label{eigen}
\]
where $\alpha =0,\cdots,  {|\rm p|}-1$ is the band index, and their eigenenergies are $\mathcal E_{0, \bm k}\leq \cdots\leq \mathcal E_{ {|\rm p|}-1, \bm k}$. The crystal momentum is restricted to $\{ | k_x|\leq \pi/a,  |k_y |\leq \pi/a \} $. Note that these states are the eigenstates of the lattice translation operators: $T_1^ {\rm p} |\psi_{\alpha \bm k}\rangle =e^{i k_x a}|\psi_{\alpha \bm k}\rangle$ and $T_2^ {\rm p} |\psi_{\alpha \bm k}\rangle =e^{i k_y a}|\psi_{\alpha \bm k}\rangle$.   

\begin{figure}[t]
\includegraphics[clip, width=\columnwidth]{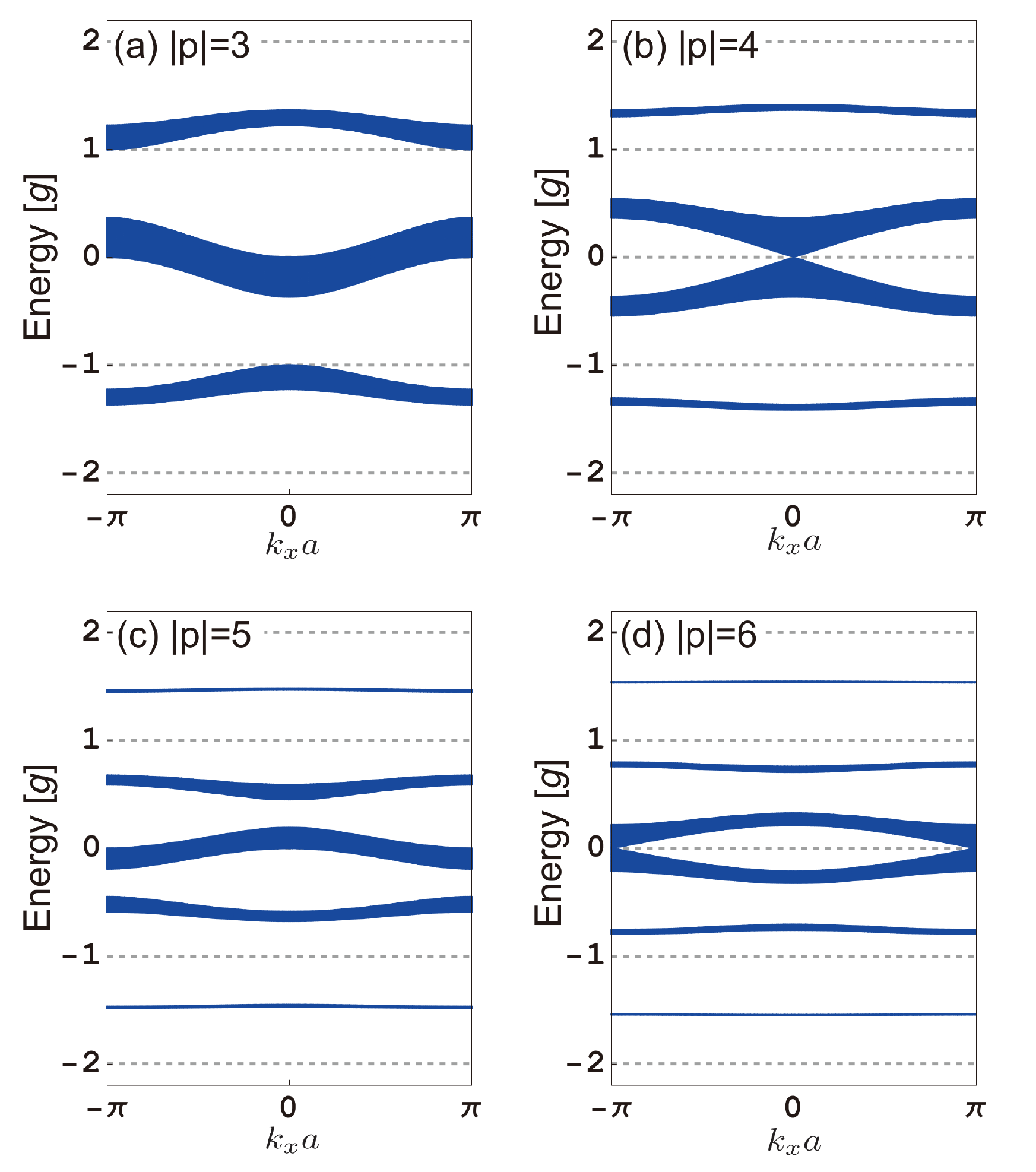}
\caption{ Energy spectrum of a single skyrmion excitation for (a) ${\rm |p|} \equiv 2S |\mathcal N|=3$, (b) ${\rm |p|}=4$, (c) ${\rm |p|}=5$, and (d) ${\rm |p|}=6$. The vertical axis represents energy in the unit of $g >0$, and the energy is measured from the energy in the continuum limit. Eigenenergies for different $k_y$ are plotted. 
}
\label{band1}
\end{figure}

\begin{figure}[t]
\includegraphics[clip, width=0.9\columnwidth]{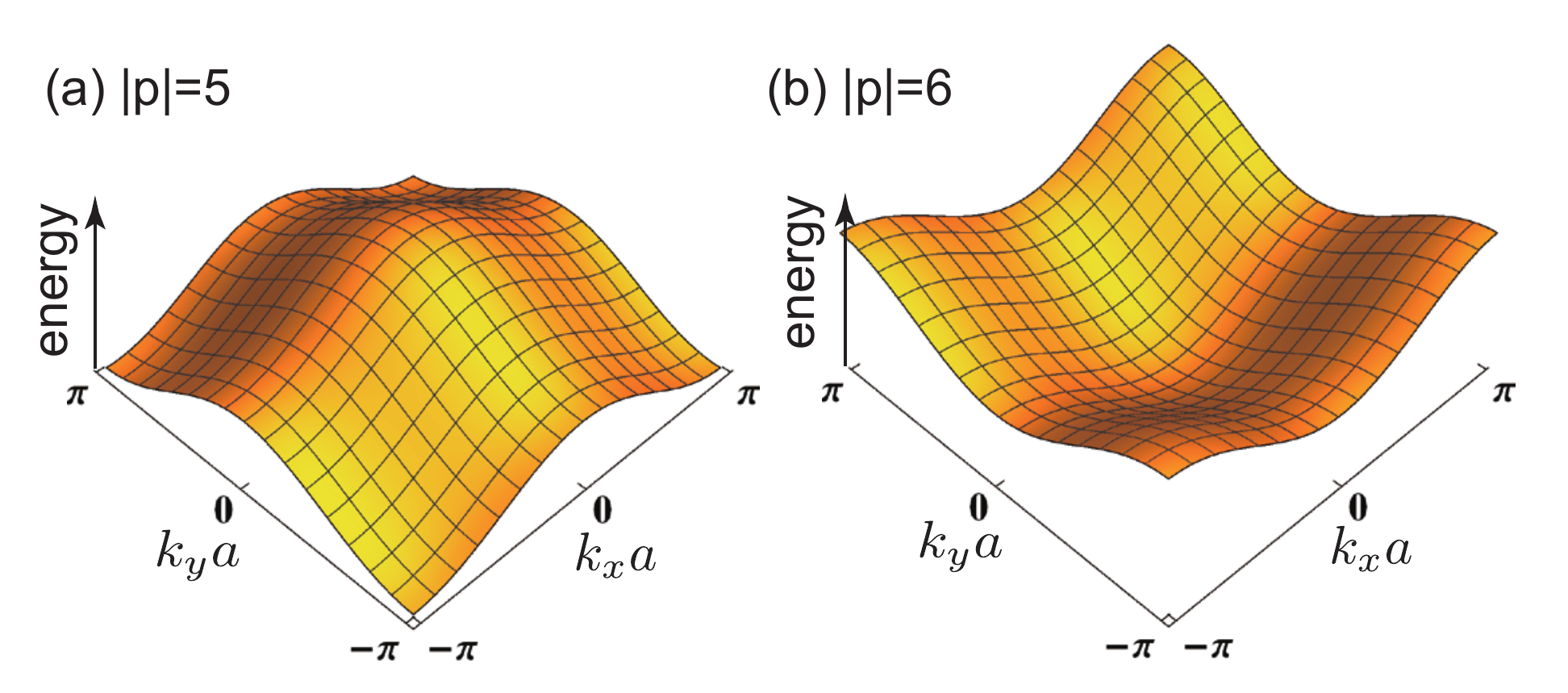}
\caption{
Band dispersion of the lowest bands for $g>0$ (a) when ${\rm p}$ is odd (${\rm |p|}=5$) and (b) when ${\rm p}$ is even (${\rm| p|}=6$).}
\label{band2}
\end{figure}

The eigenenergies of Harper's equation are well studied.\cite{Hofstadter1976} In Fig.~\ref{band1}, we show the energy spectrum of the Hamiltonian (Eq.~\eqref{harper}). Fig.~\ref{band2} shows the lowest band dispersion, which has a single minimum at $\bm k_{\text {min}} =(0,0)\ [\bm k_{\text{min}}= (\pm \pi,\pm \pi)$] when $ {\rm p}=2S\mathcal N $ is even (odd) for $g>0$. The value of $\bm k_{\text {min}}$ alters the critical behavior, which will be discussed in the next subsection. 

\section{Phase Diagram}\label{intermediate}
In this section, we study the phase diagram in the presence of interactions between skyrmions (Fig.~\ref{phase}). In the following analysis, we assume $g>0$, and we comment on the case with $g<0$ in the end of this section.  Focusing on the lowest energy band, we consider a many body action: $\mathcal S =\mathcal S_0 +\mathcal S'$ with 
\[
\mathcal S_0 = \int d\tau \left[
\sum_{\bm k}  \bar b_{\bm k} (\partial_\tau +\xi_{\bm k}) b_{\bm k}+\sum_{\bm k, \bm k', \bm q}U_{\bm k, \bm k', \bm q} \bar{b}_{\bm k+ \bm q}\bar{b}_{\bm k' -\bm q} b_{\bm k'}b_{\bm k}
\right].
\]
 $b_{\bm k} =b_{\bm k}(\tau)$ and $\bar{b}_{\bm k}= b^*_{\bm k}=\bar{b}_{\bm k}(\tau)$ are canonical Bose operators that annihilate and create the skyrmion excitations obtained in the last section. 
\[
\xi_{\bm k} = \mathcal E_{0,\bm k}+E_0 - E_{\text {FM}}
\] is the single particle energy of skyrmions measured from the FM energy, where $E_0$ is the energy of a single skyrmion in the continuum limit, and $E_{FM}$ is the energy of a FM state. 
The last term is a short range repulsive interaction, which eventually leads to the crystallization of skyrmions. $\mathcal S'$ represents the terms that do not conserve the number of skyrmions, which are allowed due to the DM interaction; it breaks the U(1) spin rotation symmetry about the applied field, and hence violates conservation of $S^z$.   Thus although a skyrmion possesses flipped spins in its core, its spin is not a good quantum number and cannot microscopically protect the skyrmion number. The processes that change the skyrmion number may be considered tunneling events which can occur due to lattice scale physics \cite{Diaz2016}.
In a high field with $E_0 \gg E_{FM}$, the density of skrymions is suppressed
\[
\langle n_{\bm k} \rangle \equiv \langle \bar{b}_{\bm k}b_{\bm k}\rangle \sim 0.
\]
In a lower field with $\xi_{\bm k_{\min}} \sim 0$, the density of skyrmions increases, with the first skyrmions entering being those with $\bm k= \bm k_{\min}$: 
\[
\langle n_{\bm k_{\text{min}}} \rangle\gg1. 
\]
The resulting quantum state is a quantum liquid of skyrmions, where skyrmions ``condense'' at $\bm k=\bm k_{\text{min}}$, and they are not spatially localized.   We note that  this condensation  differs from ideal Bose Einstein condensation since the action does not conserve skyrmion number.  

To clarify the nature of the condensation, we focus on the states around the minimum of $\xi_{\bm k}$, and define a new field $b(\bm r, \tau)\sim \eta(\bm r, \tau) e^{i \bm k_{\text{min}}\cdot\bm r }$, where $b(\bm r, \tau)$ is the inverse Fourier transform of $b_{\bm k}(\tau)$, and $\eta(\bm r, \tau)$ has small space time gradients. 
Then we obtain
\[
\mathcal S \sim  \int d\tau d^2 r
 \left[
\bar{\eta}\partial_\tau \eta +r|\eta|^2 +c_0|\bm \nabla  \eta|^2  +c_1 |\eta|^4 \right]+ \mathcal S',\label{eta_S}
\]
where $r=\xi_{\bm k_{\min}} $, $c_0 =\frac{1}{2}\partial^2_{k_\mu} \xi_{\bm k_{\rm min}}$, and $c_1$ is a constant given by the interaction.   

For odd ${\rm p}=2S\mathcal N $, there is a continuous phase transition. Since the energy minimum is at $\bm k_{\min} = (\pi ,\pi )$, the odd order terms of $\eta, \bar \eta$ are forbidden in $\mathcal S'$ by translational symmetry/momentum conservation: 
\[
\mathcal S^\prime \sim \int d\tau d^2  r \left[  f^{(2)} \eta \eta  +(f^{(4a)} \eta\eta \eta\eta+\cdots)  +\text{c.c.}
\right].
\]
To obtain the critical theory, we define $\eta(\bm r, \tau) =\varphi_R(\bm r,\tau)+i\varphi_I(\bm r,\tau)$. 
Up to quadratic order, the total action is given by  
\[
\mathcal S &\sim \int d\tau d^2 r \left[
 -2 \varphi_I (i \partial_\tau +F_1)\varphi_R+\varphi_R(r+F_0-c_0 \partial_\mu^2)\varphi_R\nonumber \right.\\
&\left.
\hspace{60pt}+\varphi_I(r-F_0-c_0 \partial_\mu^2)\varphi_I
\right].
\]
We have defined $F_1 =2 {\rm Im} f^{(2)}$ and $F_0 =2 {\rm Re} f^{(2)} (> 0)$, the sign of which can be arbitrarily chosen by redefining $\eta$. 
In the critical region where the single particle energy gap of $\varphi_I$ becomes small $(r-F_0 \sim 0)$, we can integrate out $\varphi_R$, which has the energy gap $\sim 2F_0$. Then, the critical behavior is described by the following action: 
\[
\mathcal S&\sim  \int  d\tau d^2r \left(
\frac{1}{2}\left((\partial_\tau \varphi_{I})^2+r' \varphi_{I}^2 +v(\partial_\mu \varphi_{I})^2 \right)  +\frac{u }{4!} \varphi_{I}^4 
\right), \label{critical}
\]
where $v, u$ and $r'= (r+F_0)(r-F_0)-F_1^2$ are constants, and $\varphi_I$ is redefined to absorb some constants. (See Appendix~\ref{a:critical} for the detail.)
 This is the standard $\varphi^4$ action describing the Ising phase transition, e.g. in the three-dimensional ferromagnetic classical Ising model.  The zero temperature transition in $2+1$ space-time dimensions has quantum critical behavior in this universality class.  The phase transition is described by the order parameter $\langle \varphi_I \rangle$ that breaks the symmetry $\varphi_I \rightarrow - \varphi_I$. Thus, the condensation of skyrmions is accompanied by the phase transition at $B^*$ with the critical behavior given by Eq.~\eqref{critical}. 

On the other hand, for even  ${\rm p}$, the energy minimum is at $\bm k_{\min} = (0,0)$, and we have    
\[
\mathcal S^\prime &\sim \int d\tau d^2 r \left[  f^{(1)}\eta + f^{(2)} \eta \eta +\cdots  +\text{c.c.}
\right].
\]
The first-order terms,  $f^{(1)} \eta +f^{(1) *} \bar{\eta}$ give rise to $-h \varphi_I $ in the final action [Eq.~\eqref{critical}] with a constant $h$. It results in a crossover, which is the same as the Ising model under a magnetic field.

In this case, there is no true quantum phase transition corresponding to the condensation of skyrmions.    To understand this, note that, both for even and odd p, some virtual skyrmions are present in the ground state at all fields, even above the na\"ive condensation point.  In the case of even p, the real (not virtual) skyrmion state with minimum energy has the same quantum numbers as the ground state (e.g. momentum zero), and so a level crossing of this excited state with the ground state cannot occur due to level repulsion.  This explains the absence of a condensation phase transition for even p.  However, at lower fields, a transition to a SkX state still occurs. 

Next we discuss this phase transition from the quantum liquid state to the SkX phase, which occurs at the even lower field ($B=B_c$). Let us define the typical magnitude of the two-body repulsive interaction $\bar U(n) $, which depends on the distance between skyrmions $\sim n^{-1/2}$.   
The band width $W$ of the skyrmion excitation is bounded by $g$, where $g$ is estimated in Eq.~\eqref{g_order}. 
 For $ W \gg \bar U(n) $, the kinetic energy is dominant, so that the skyrmions have itinerant properties (quantum liquid). When the density becomes large enough such that $ W \ll \bar U(n) $, the repulsive interaction becomes dominant, and skyrmions form a density wave, i.e. a SkX, to minimize the interaction. The density-density correlation has the Fourier expansion 
\[
\langle n (\bm r)n (\bm 0)\rangle =\text{const.} + {\rm Re }\sum_i A_i e^{i \bm Q_i \cdot \bm r} +\cdots ,
\]  
where $n(\bm r)$ is the local density for skyrmions, 
and $\bm Q_i $ are incommensurate wave vectors determined by the structure of the interaction.   The amplitudes $A_i$ are order parameters for the SkX phase, which breaks lattice translational invariance. The critical field $B_c$ is determined by the critical density $n_c$ that satisfies $\bar U(n_c) \sim W$. Thus, for smaller bandwidth, the region of the quantum liquid phase becomes narrower. 

Finally, we also comment on the case with $g<0$. In this case,  the energy minimum of a skyrmion is at $k_{\rm min}=(0,0)$ regardless of $2S\mathcal N$, and it results in the crossover between a quantum liquid and a FM state.   
\section{Inelastic Neutron Scattering} \label{neutron}

In this section, we discuss how single skyrmion excitations can be observed in the FM  state.  We first introduce an approximated wave function for a single skyrmion excitation: 
\[
b^\dag_{\bm k}|{\rm FM} \rangle \sim \frac{1}{N}\sum_s e^{-i \bm k\cdot \bm R_s} \psi^{\dag}_{\bm R_s}|{\rm FM} \rangle,
\]
where $\bm R_s=(a/2,a/2)+(s_x a,s_y a)$ with $s_x, s_y \in \mathbb Z$, 
$|{\rm FM} \rangle$ denotes a FM state, $b_{\bm k}$ is the operator for a skyrmion excitation, and 
\[
\psi^{\dag}_{\bm R_s}|{\rm FM} \rangle = \otimes_i | \bm n_{sk} (\bm r_i-\bm R_s) \rangle, 
\]  
is a state with a skyrmion excited at $\bm R_s$.

\begin{figure}[t]
\includegraphics[clip, width=0.4\columnwidth]{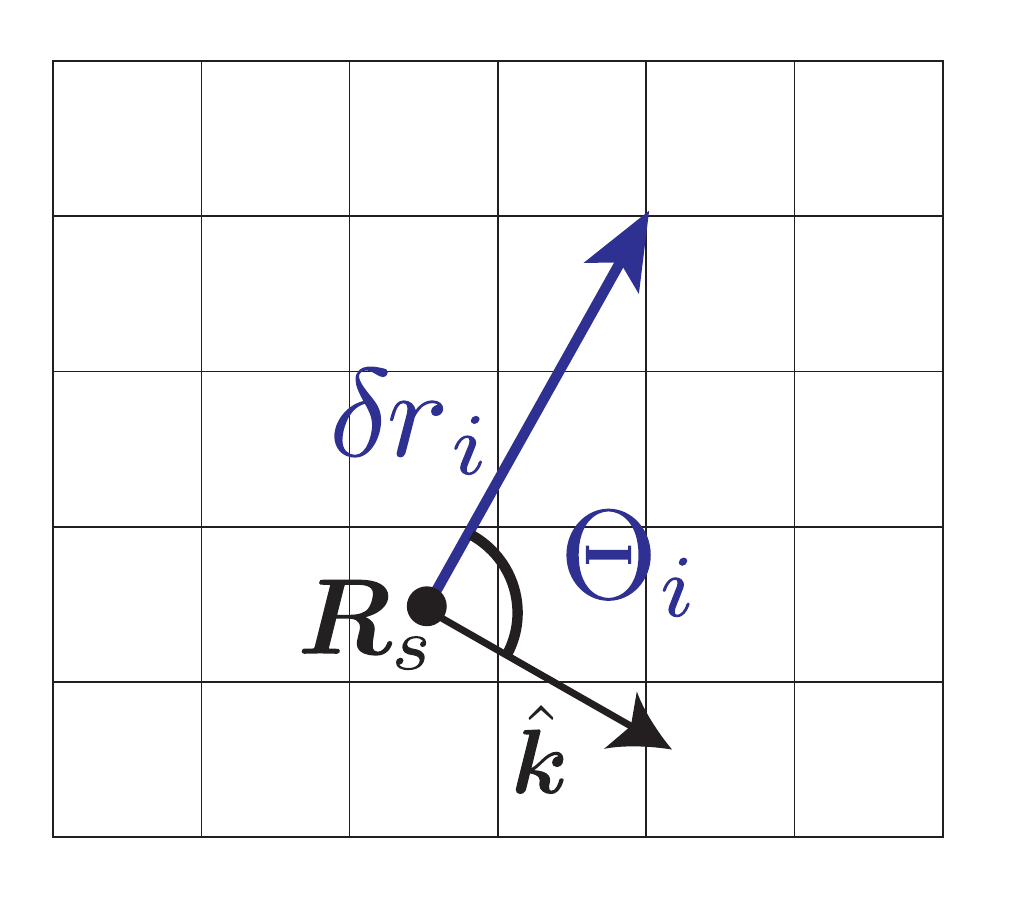}
\caption{
Polar coordinate for a skrymion configuration. $\bm R_s$ is the center of the skyrmion. 
}
\label{polar}
\end{figure}

Recall that a spin coherent state at each site can be represented by
\[
|\bm n_i \rangle &=e^{i S_z \phi}e^{i S_y \theta}e^{i S_z \chi}|S,S\rangle,\\
&=\left(\cos \frac{\theta_i}{2}\right)^{2S}\sum_{m=0}^{2S}\frac{1}{m!} \left(\tan \frac{\theta_i}{2}\right)^m e^{i m\phi _i}(\hat{S}^-_i)^m |S,S\rangle, \label{coherent}
\]
where we have chosen the gauge as $\chi = -\phi$, $\bm n_i =(\sin \theta_i \cos \phi_i,\sin \theta_i \sin \phi_i, \cos \theta_i)$ and $|S,S\rangle $ is the maximally polarized state of a single spin. 
The skyrmion configuration  $ \bm n_i  = \bm n_{sk}(\bm r_i-\bm R_s \equiv \delta \bm r_i) $ can be simply represented with a polar coordinate $\delta \bm r= (\delta r_i, \Theta_i)$ centered at $\bm R_s$ (Fig.~\ref{polar}):  
\[
\theta_i &=\theta(\delta r_i), \label{config1}\\
\phi_i &=\Theta_i + \alpha,\label{config2}
\] 
where $\theta(0)=\pi$ and $\theta(\delta r \gg L_s)\sim 0 $. Here we consider a configuration stabilized by the DM interaction, i.e. $\mathcal N=-1$ with the fixed ``helicity'' $\alpha$. 

As is clear from Eq.~\eqref{coherent}, a skyrmion state is a superposition of states with different numbers, $m_{tot}$ of bound magnons: $m_{tot} \equiv \sum_i(S-S_i^z)$, which includes a state with $m_{tot}=1$. This state can be captured by inelastic neutron-scattering measurements, which measures the dynamical spin correlation 
\[
{\rm Im}\ \chi^{+-}(\bm Q, \omega)&={\rm Im} \left [ 
i \int_0^\infty dt e^{i \omega t}\langle \hat S^{+}_{\bm Q}(t)\hat S^{-}_{-\bm Q}(0) \rangle
\right].\]

Let us estimate the magnitude of the signal of a single skyrmion excitation:   
\[{\rm Im}\ \chi^{+-}(\bm Q, \omega)&\sim \pi \sum_{\bm k} | \langle {\rm FM}| b_{\bm k}\hat{S}_{-\bm Q}^- |{\rm FM} \rangle |^2 \delta(\omega-\xi_{\bm k}). 
\]
Using the explicit configuration $ \bm n_i$ in Eqs.~\eqref{config1} and \eqref{config2}, we obtain   
 \[&| \langle {\rm FM} | b_{\bm k} \hat{S}_{-\bm Q}^- |{\rm FM} \rangle |  \nonumber\\
&=\frac{S}{N} \left(\prod_{j \in A} \cos\frac{\theta(\delta r_j) }{2}\right)^{2S}\left| \sum_{i\in A}  \tan{\frac{\theta(\delta r_i)}{2}} e^{i ( Q \delta r_i \cos \Theta_i -\Theta_i  )} \right|\delta_{\bm Q, \bm k},
\label{matrix}\]
where $A $ is the region such that $\theta (\delta r_{i\in A}) \neq 0$; the region where a skyrmion spreads over. The number of sites in $A$ is $\sim (L_s /a)^2 $. 

For $L_s^{-1}\ll Q \ll a^{-1}$, we can further estimate Eq.~\eqref{matrix} with a continuous approximation 
\[{\rm Im}\ \chi^{+-}(\bm Q, \omega)\sim
 \frac{1}{a^2 Q^2}  \exp\left( {-(\ln 2) S(L_s/a)^2}\right)\delta(\omega-\xi_{\bm Q}),\]
where $\xi_Q$ is the excitation energy of a skyrmion defined in the last section.
 Although the intensity remains weak, it should be distinct from the signal of a magnon; skyrmion excitations have multiple bands, and the dispersion is much smaller. 
 
\section{Summary and discussion}\label{conclusion}

In this paper, we have studied the quantum description of magnetic skyrmions in a two-dimensional chiral magnet. Such quantum mechanical properties may appear for small skyrmions, especially in the region close to a SkX phase. We have shown that a quantum liquid phase can appear as an intermediate phase. 

A basic result is that the well-known classical Magnus force dynamics of skyrmions,\cite{Stone1996, Jonietz2010,Yu2012a}\ extends to the quantum level and dominates the band structure of skyrmion states, making them quantitatively very different from the usual magnon band(s) of a ferromagnet.  While this is not surprising from a semi-classical point of view, it may raise flags for a many-body quantum physicist.  The skyrmion is a local excitation (i.e. it does not affect spins far from its core) of a ferromagnetic state, which has only short-range entanglement: it is essentially a product state of up spins.  The microscopic spin Hamiltonian for a chiral magnet is local and the spins are neutral and do not transform under any gauge symmetry.  In general, we expect that a short-range entangled state of such a Hamiltonian would possess only neutral quasiparticles which could not experience any orbital magnetic fields (i.e. they do not couple to any gauge fields).  More formally, a charged particle in a magnetic field transforms under translations differently from a neutral particle.  The latter transforms under simple coordinate transformations only, while for the former particle, a coordinate transformation must be accompanied by a gauge transformation as the vector potential is not translationally invariant: such transformations are known as magnetic translations.  It would be exceedingly strange (we believe impossible) for a skyrmion in a ferromagnet to truly exhibit such a gauge structure.  In our calculations here we found a resolution to this dilemma.  Specifically, taking into account even weak lattice effects, which are necessary for a finite quantum theory, the effective number of flux quanta per unit cell is found to be an integer.  Under this condition magnetic and ordinary translations coincide.  Consequently, the bands of skyrmion states are not {\sl qualitatively} distinct from those of magnons, although we find that quantitatively they are very different and should be easily differentiated in experiment. 

In our study, we have not included a mass term $\sim (\dot X^2+\dot Y^2)$ in the effective action. Microscopically it can arise from coupling to some gapped modes, such as the deformation of a skyrmion.~\cite{Makhfudz2012} The mass term makes skyrmions form Landau bands, which are flat without periodic potentials. Therefore the interaction between skyrmions immediately forces them to form a crystal; this is consistent with the classical models. However, once the cosine potential is considered, it recovers the subbands structure with dispersion.~\cite{Usov1988, Pfannkuche1992} From this point of view, our study corresponds to projection onto the lowest Landau level.  

Regarding experiments, hexagonal Fe film on the Ir $(111)$ surface~\cite{Heinze2011} and thin films of Fe$_{0.5}$Co$_{0.5}$Si and MnSi grown along the $\langle 111\rangle$ direction have lattice structures effectively modeled by triangular lattices. We, however, note that the triangular lattice also gives qualitatively the same results; the band splits into $|{\rm p}|=2S|\mathcal N|$ sub-bands, and a crossover to the quantum phase for even ${\rm p}$, and a quantum phase transition for odd ${\rm p}$. 

We note that the magnetization of quantum skyrmions is a fluctuating quantum field, but it would be possible to calculate the average magnetization profile of the quantum skyrmion to compare it with classical configurations of a skyrmion as a further analysis. 

Finally, we comment on skyrmions in frustrated spin systems. Theoretically it has been shown that SkX phases appear on a triangular lattice with frustrated interactions, which preserves spin U(1) symmetry along the field\cite{Okubo2012,Leonov2015, Lin2015}. We showed that the quantum states of skyrmions in chiral magnets have crystal momentum as a good quantum number [Eq.~\eqref{eigen}]. On the other hand, skyrmions in frustrated magnets have additional quantum numbers: the number of bound magnons $m_{tot}$ and the discrete angular momentum related to the helicity of a skyrmion. The magnon number $m_{tot}$ is related to the size of a classical configuration of a skyrmion. It would be interesting to investigate the quantum phase and critical theory for such skyrmions. We leave the detailed discussion for future work.  
 \begin{acknowledgements}
We thank Jiangpeng Liu, Satoshi Fujimoto, and Masatoshi Imada for fruitful discussions. R.T. was supported by the Japan Society for the Promotion of Science (JSPS) Fellowship for Young Scientists. H.I. was supported by the JSPS Postdoctoral Fellowship for Research Abroad.  L.B. was supported by the NSF Materials Theory program Grant No. DMR1506119.  We benefited from the facilities of the Kavli Institute for Theoretical Physics, and so were supported in part by NSF Grant No. NSF PHY1125915.\end{acknowledgements}

\appendix
\section{Effective action}\label{ap:eff_S}
We first show that the periodic potential in Eq.~\eqref{effS} is obtained from $H(\{\bm n_i\})$ in Eq.~\eqref{action}. Let us define the ``energy density'' $h (\bm r)$ as
\[
&{\rm H}(\{\bm n_i\}) \equiv a^2 \sum_{ i} h \left(\bm r_i- \bm R \right),\\
&h (\bm r-\bm R) =\frac{JS^2}{a^2} - \frac{JS^2 }{2} \partial_ \mu \bm n\cdot \partial_\mu \bm n + \frac{DS^2}{a}  \sum_{\mu=x,y}\hat{\bm e}_\mu \cdot \left(\bm n\times \partial_\mu \bm n\right)\nonumber\\
&\hspace{45pt}- \frac{BS}{a^2} n^z -\frac{K S  \left(S-\frac{1}{2}\right) }{a^2}(n^{z})^2+  O\left(\frac{Ja^2}{L_s^4} \right),
\]
where  $\bm r_i =(i_x a,i_y a)$ denotes the position of a site $i$, and $\bm n$ abbreviates $\bm n_{sk} (\bm r-\bm R)$.

{We consider skyrmions whose energy density $h(\bm r)$ has the maximum at the center $\bm r= \bm 0$, the center of the skyrmion.} The characteristic length of $h(\bm r)$ is given by the skyrmion configuration as $L_s  (\gg a)$. We then expand the discrete summation  
with the Poisson summation formula $\sum_{ i}\delta^{(2)}(\bm r- \bm r_i) =a^{-2} \sum_{l_x, l_y}  e^{-\frac{2\pi i}{a} \bm l \cdot \bm r }$ as 
\[
 {\rm H}(\{\bm n_i\})
&\sim \int d^2 r  h(\bm r-\bm R) \nonumber\\
&\hspace{15pt}+2\int d\bm r \left( \cos \left(\frac{2\pi  x}{a}\right)+ \cos \left(\frac{2\pi  y}{a}\right) \right) h(\bm r-\bm R), \label{ham1}\\
&\sim E_{0} +g\left( \cos \left(\frac{2\pi X}{a}\right) +\cos \left(\frac{2\pi Y}{a} \right) \right), \label{lattice}
\]
$E_0$ is the energy in the continuous limit, which is constant for $\bm R$. The second term in Eq.~\eqref{ham1} gives the periodic potential for $\bm R$ with the coefficient 
\[
g&\sim  \frac{8\pi\   h(\bm 0) ^2}{|\partial^2_{ \mu}  h(\bm 0)| } \exp{\left( -\frac{4\pi ^2  h (\bm 0)  }{ a^2|\partial^2_{ \mu}  h(\bm 0)|  }\right)  }, \\
&= \frac{2C}{\pi}L_s^2h(\bm 0)  \exp\left(-C \frac{L_s^2 }{a^2} \right) . \label{g_order}
\]
Here we have assumed $h(\bm r)$ has rotational symmetry, and expanded $h(\bm r)=\exp(\log h(\bm r)) \sim h( 0) \exp\left(\frac{1}{2h(0)}  (x^2 \partial^2_{x}  h(0)+y^2 \partial^2_{y}  h(0) )\right) $. We define a dimensionless constant $C \equiv 4\pi ^2  h (\bm 0)  /(L_s^2|\partial^2_{\mu}  h(\bm 0)|) \sim  O (1)$. The higher harmonics, which are dropped in Eq.~\eqref{ham1}, gives different periodic potentials, but their strength is higher order of $ \exp{( -CL_s^2/a^2) }$. In Eq.~\eqref{g_order}, we denote $h(\bm 0) $ by ${\varepsilon}$.  

Next we discuss the Berry phase terms. As in Eq.~\eqref{ham1}, we can expand the discrete summation over sites. To the lowest order, we obtain 

\[i S \int d\tau  \sum_i \dot{\bm n}_i \cdot \bm A(\bm n_i )&\sim
\frac{iS}{a^2} \int d\tau \int d \bm r\dot{\bm n}  \cdot \bm A(\bm n ), \label{Berry}\\
&=\frac{2\pi i  S \mathcal N}{a^2}\int d\tau  (Y\dot{X}-X\dot{Y}),
\]
 up to the surface terms in the time integral~\cite{Stone1996}.  
The rest of the terms in the harmonic expansion are smaller than Eq.~\eqref{lattice} because of the time derivative, and thus we take only the lowest order, i.e. the continuous limit.

\section{Critical Theory}\label{a:critical}
We show the derivation of the effective action in the critical region  for even $2S\mathcal N$ and odd $2S\mathcal N$ (Eq.~\eqref{critical}).
We consider the action up to quadratic order
\[
\mathcal S &\sim  \int d\tau d\bm r \left[
\bar{\eta}\partial_\tau \eta +r|\eta|^2 +c_0|\bm \nabla  \eta|^2 \right. \nonumber\\
&\left.\hspace{50pt}
+f^{(1)} \eta +f^{(1) *} \bar\eta+f^{(2)} \eta \eta+f^{(2) *} \bar{\eta}\ \bar{ \eta} 
\right],
\] 
Note that $f^{(1)}=0$ for odd $2S\mathcal N$.  
We now define $D_0 =2 {\rm Re} f^{(1)} $ and $D_1 = 2{\rm Im} f^{(1)}$, $F_0 =2 {\rm Re} f^{(2)} > 0$, $F_1 =2 {\rm Im} f^{(2)} $, and $\eta(\bm r, \tau) =\varphi_R(\bm r,\tau)+i\varphi_I(\bm r,\tau)$, and obtain
\[
\mathcal S &= \int d\tau d\bm r \left[
 -2i \varphi_I \partial_\tau \varphi_R+\varphi_R(r-c_0 \partial_\mu^2+F_0)\varphi_R\nonumber\right. \\
&\hspace{20pt}\left. +\varphi_I(r-c_0 \partial_\mu^2-F_0)\varphi_I-2F_1 \varphi_R\varphi_I
+D_0 \varphi_R- D_1 \varphi_I  
\right].
\]
For $r \sim F_0$, we integrate out $\varphi_R $, and obtain 
\[
\mathcal S&=  \int d \bm r d\tau \left(
\frac{1}{2}\left((\partial_\tau \varphi_{I})^2+r' \varphi_{I}^2 +v(\partial_\mu \varphi_{I})^2 \right)  -h \varphi_I+\frac{u }{4!} \varphi_{I}^4 
\right),  \label{final}
\]
where we redefine $\varphi\rightarrow \sqrt{\frac{2}{r+F_0}} \varphi_I$, and 
\[
r'&=(r+F_0)(r-F_0)-F_1^2, \\
v&=c_0 \left(r+F_0+\frac{F_1^2}{r+F_0}\right), \\
h&=\sqrt{ \frac{ r+F_0}{2}} \left(D_1 -\frac{F_1D_0 }{r+F_0  } \right).
\]
We have recovered the interaction term in Eq.~\eqref{final}. 
For odd 2$S \mathcal N$, we note that $h=0$.

\bibliographystyle{apsrev4-1}

\end{document}